\documentclass[slac_one]{revtex4}
\usepackage{graphicx}
\usepackage{fancyhdr}
\usepackage{epstopdf}
\newcommand{\etal}{{\em et. al.}}

\begin{document}

\title{Rare Decays of $B$ and $D$ Hadrons at CDF} %% Paper title goes here


\author{R.F.~Harr}
\affiliation{Wayne State University, Detroit MI 48202, USA \newline
	for the CDF Collaboration}

\begin{abstract}
Decays that are highly suppressed in the standard model
%, but otherwise conserve basic quantities, 
are excellent places to search for effects of new physics.
Decays mediated by flavor-changing neutral currents are forbidden at tree level in the SM, and are often further suppressed by helicity and the Glashow-Iliopoulos-Maiani  mechanism.
Exclusive final states with charged lepton pairs are a particular strength of the CDF experiment due to the large bottom and charm production cross section and the ability to efficiently separate signal from background.
CDF has searched for and set the world's best limits on the rare flavor-changing neutral current decays
$B^0_{(s)} \rightarrow \mu^+\mu^-$,
$B^0_{(s)} \rightarrow e^+ e^-$, and
$D^0 \rightarrow \mu^+\mu^-$, and the lepton flavor violating decay
$B^0_{(s)} \rightarrow e^{\pm} \mu^{\mp}$.
\end{abstract}

\maketitle

\thispagestyle{fancy}


\section{INTRODUCTION}

\begin{figure*}[htbp]
\centering
\includegraphics[width=120mm]{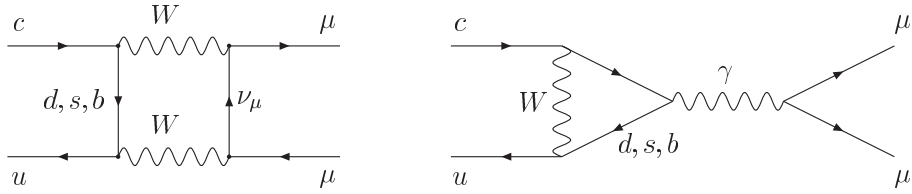}
\caption{Short distance Feynman diagrams for $D^0 \rightarrow \mu^+\mu^-$.} 
\label{fig:feyndiag}
\end{figure*}

Flavor-changing neutral current (FCNC) decays are forbidden at tree level in the standard model (SM).
They can proceed through loop processes -- an example is displayed in Fig.~\ref{fig:feyndiag} -- at a considerably suppressed rate.
The modes under study have additional helicity and GIM suppression, and have SM branching fractions beyond the reach of current experiments.
The largest predicted branching fraction is about $4\times10^{-9}$ for the decay $B^0_s \rightarrow \mu^+\mu^-$, the others being more than an order of magnitude smaller.

A primary interest in these modes is as probes of new physics (NP).
NP can enter either at tree level, or through loop processes, and can increase the branching fraction by orders of magnitude.
%, usually up to the present experimental limit.
These modes provide the best limits on a number of NP scenarios.

Lepton-flavor violating (LFV) decays are strictly forbidden in the SM, so observation is a clear sign of NP.
(The LFV in neutrino oscillations plays an insignificant role in the decays studied here.)

CDF searched for the decays $B^0_{(s)} \rightarrow \mu^+\mu^-$, $B^0_{(s)} \rightarrow e^+ e^-$, $B^0_{(s)} \rightarrow e^{\pm} \mu^{\mp}$, and $D^0 \rightarrow \mu^+\mu^-$.
(Inclusion of change conjugate modes is assumed throughout.)
The searches share many features.
The search is performed relative to a normalization mode chosen for its similarity to the search mode.
This avoids the difficulty of determining absolute efficiencies, and largely cancels many systematics.

The selection criteria for the search mode are optimized in an unbiased manner, with the events in the search window blinded.
A Monte Carlo with a detailed detector simulation is used to simulate signal events.
Background is estimated from sideband data, data before lepton identification, and Monte Carlo when appropriate.
The production, acceptance, and reconstruction efficiency of the search mode are determined relative to the normalization mode.
With the selection optimized, the data is unblinded.
None of the investigated decays showed a significant signal after unblinding, so branching fraction limits are determined, for example, as:

\begin{equation}\label{eq:limits}
{\mathcal B}^{lim}(B^0_s \rightarrow e^+e^-) = \left(\frac{N^{lim}_{e^+e^-}}{N_{K^+\pi^-}}\right) \left(\frac{\epsilon_{K^+\pi^-}}{\epsilon_{e^+e^-}}\right) \left(\frac{f_d}{f_s}\right) {\mathcal B}(B^0_d \rightarrow K^+\pi^-)
\end{equation}
where the superscript $lim$ indicates that the corresponding quantity is a credibility level limit, $N_X$ is the number of decays observed to final state $X$, the ratio of $\epsilon$'s is the ratio of acceptance and efficiency between the search and normalization final states, ${\mathcal B}(B\rightarrow X)$ is the branching fraction of $B$ to final state $X$, and, because the search mode is for $B_s$ while the normalization mode is for $B_d$, the ratio of fragmentation fractions $f_d/f_s$ is needed.
Equation~\ref{eq:limits} is for $B^0_s \rightarrow e^+e^-$, and the obvious substitutions yield the expressions for the other modes.

In the following section we present some of the specific challenges of each analysis, then summarize the results and discuss limits on NP that can be derived.

\section{THE ANALYSES}
%\subsection{THE DECAY $B^0_{(s)} \rightarrow \mu^+\mu^-$}

The $B^0_{(s)} \rightarrow \mu^+\mu^-$ analysis for $2\,\mathrm{fb}^{-1}$ of integrated lumninosity is published \cite{bmm} and will be omitted in this proceeding in the interest of devoting more space to the unpublished results.
The limits are displayed in Table~\ref{limits}.

%\subsection{THE DECAY $B^0_{(s)} \rightarrow e^{\pm}\mu^{\mp}$}

%\subsection{THE DECAY $B^0_{(s)} \rightarrow e^+e^-$}

\begin{figure*}[t]
\centering
\includegraphics[width=120mm]{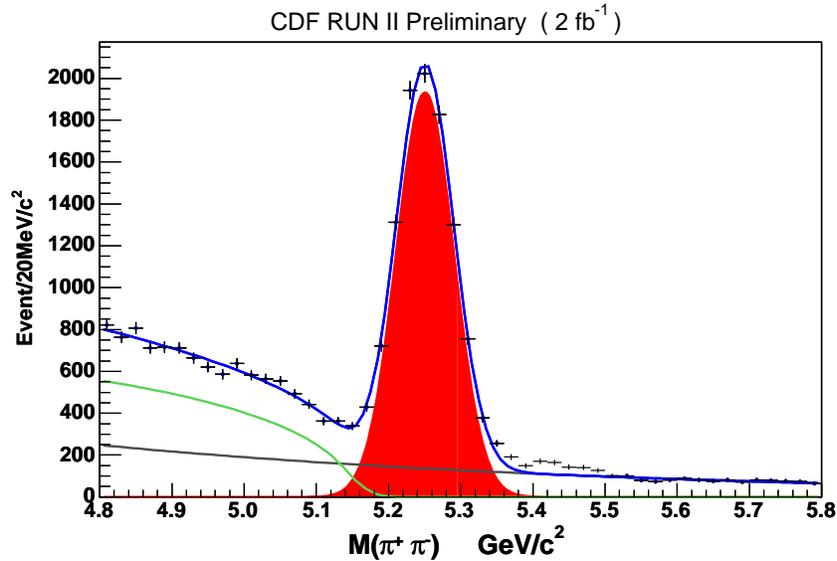}
\caption{Two track $\pi\pi$ mass plot.
The peak in the mass plot represents $9648.4\pm224.7$ $B^0_{d,s} \rightarrow hh'$ decays of which $6387.0\pm214.4$ are identified as $B^0_d \rightarrow K^+\pi^-$.} 
\label{fig:Mpipi}
\end{figure*}

The $B^0_{(s)} \rightarrow e^{\pm}\mu^{\mp}$  and $B^0_{(s)} \rightarrow e^+e^-$ decays are normalized to the charmless $B^0_d \rightarrow K^+\pi^-$ decay \cite{bemee}.
The number of $K^+\pi^-$ decays is extracted from data collected by a two-track, separated vertex trigger (TTT) from $2\,\mathrm{fb}^{-1}$ of integrated luminosity.
This trigger selects events with two oppositely charged tracks with transverse momenta above $2\,\mathrm{GeV}/c$, impact parameters between $140\,\mu\mathrm{m}$ and $1000\,\mu\mathrm{m}$ to the beam spot, and a transverse decay length greater than $200\,\mu\mathrm{m}$.
The decays $B^0_{(s,d)} \rightarrow \pi^+\pi^-, K^+\pi^-, K^+K^-$ reconstructed from the trigger track-pairs are indistinguishable in the invariant mass plot (Fig.~\ref{fig:Mpipi}), and are separated on a statistical basis as described in \cite{bpipi}.

\begin{figure*}[htbp]
\centering
\includegraphics[width=80mm]{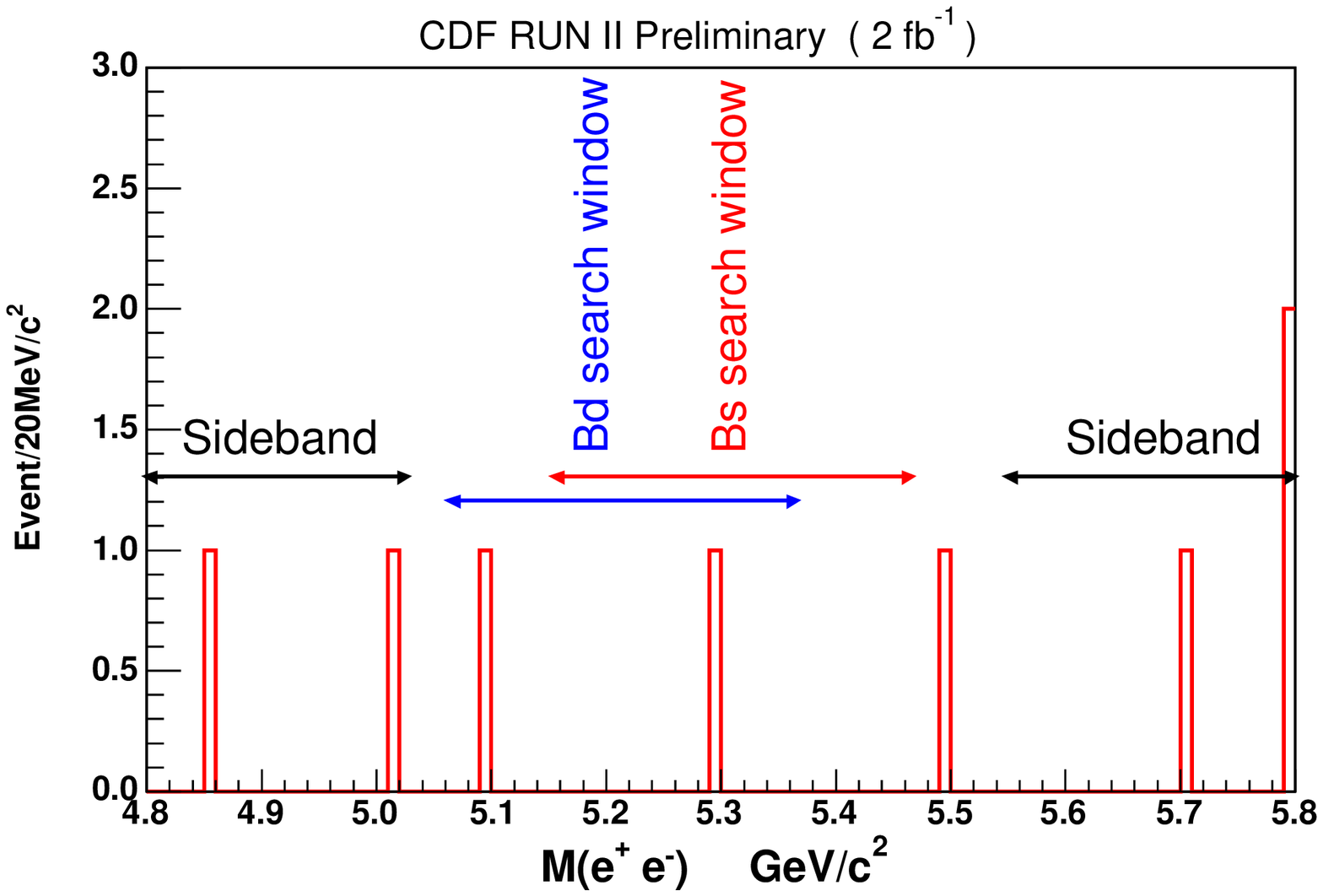}
\includegraphics[width=80mm]{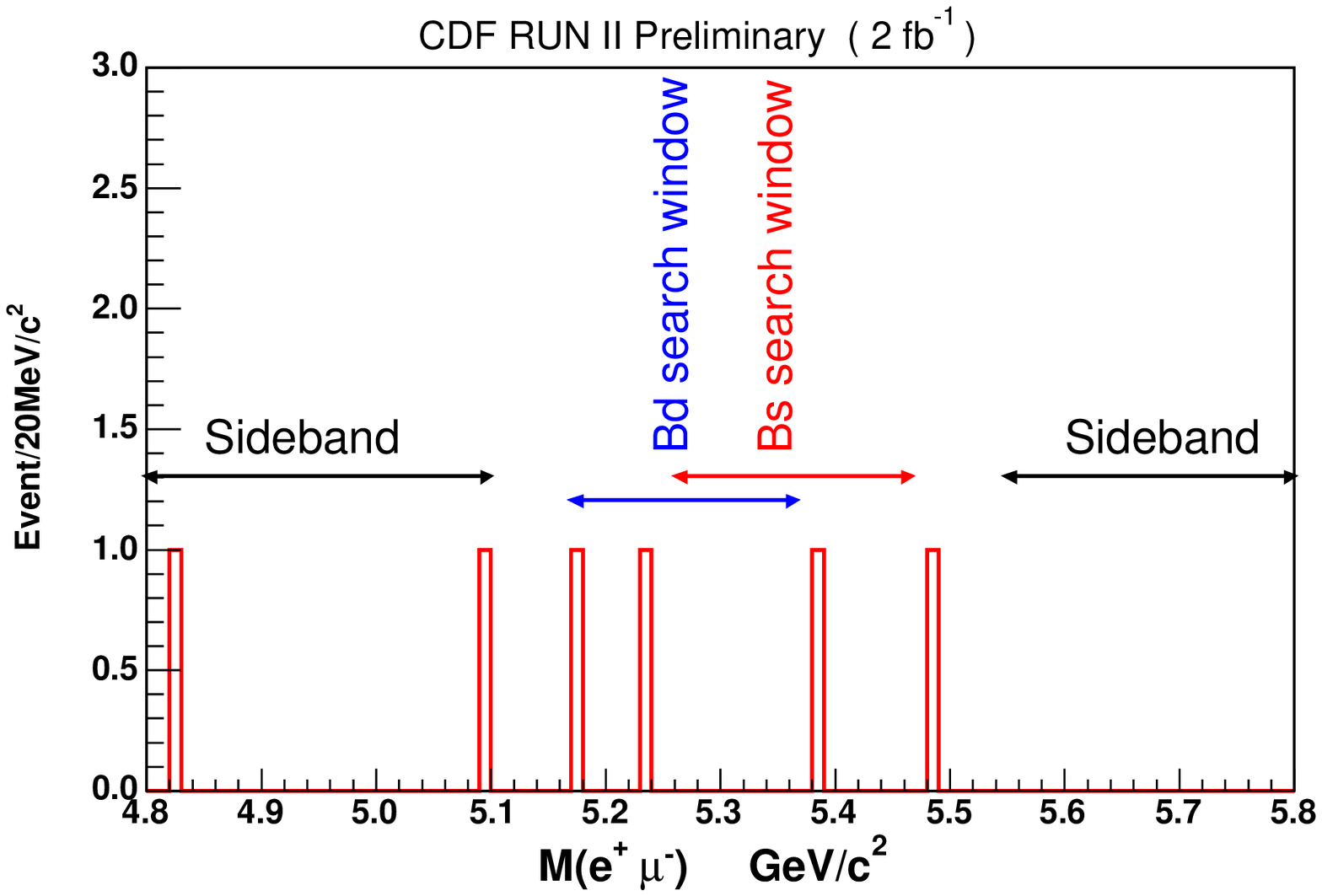}
\caption{The unblinded mass distributions after applying all selection requirements for $B^0_{(s)} \rightarrow e^+e^-$ (left) and $B^0_{(s)} \rightarrow e^{\pm}\mu^{\mp}$ (right).
The number of observed events is consistent with the background estimate.}
\label{fig:beeemuMass}
\end{figure*}

The $e\mu$ and $ee$ decays are extracted from this same data by reconstructing the trigger track-pairs as either $e^{\pm}\mu^{\mp}$ or $e^+e^-$ and then requiring the tracks be matched to hits in the muon detectors, or to an energy cluster in the electromagnetic calorimeter.
Additional background suppression for electrons is provided by requiring an ionization energy loss, $dE/dx$, in the outer tracker inconsistent with hadrons.
For electrons, bremstrahlung in the detector can result in significant energy loss producing a tail in the reconstructed invariant mass distribution. 
To maintain good acceptance, a generous search window is used, $\pm3\sigma$ around the nominal $B_s$ or $B_d$ mass for the $e\mu$ modes, and $(-6\sigma, +3\sigma)$ for the $ee$ modes.
The backgrounds are primarily combinatorial and from misidentified $B^0 \rightarrow hh'$ decays.
The unblinded mass plots (Fig.~\ref{fig:beeemuMass}) reveal two events in the $B^0_d \rightarrow e^{\pm}\mu^{\mp}$  and $e^+e^-$ mass windows, one event in the $B^0_s \rightarrow e^{\pm}\mu^{\mp}$  and $e^+e^-$ mass windows.
The $90\%$ and $95\%$ Bayesian credibility level limits determined for these modes is displayed in Table~\ref{limits}.

%\subsection{THE DECAY $D^0 \rightarrow \mu^+\mu^-$}

\begin{figure*}[htbp]
\centering
\includegraphics[width=55mm]{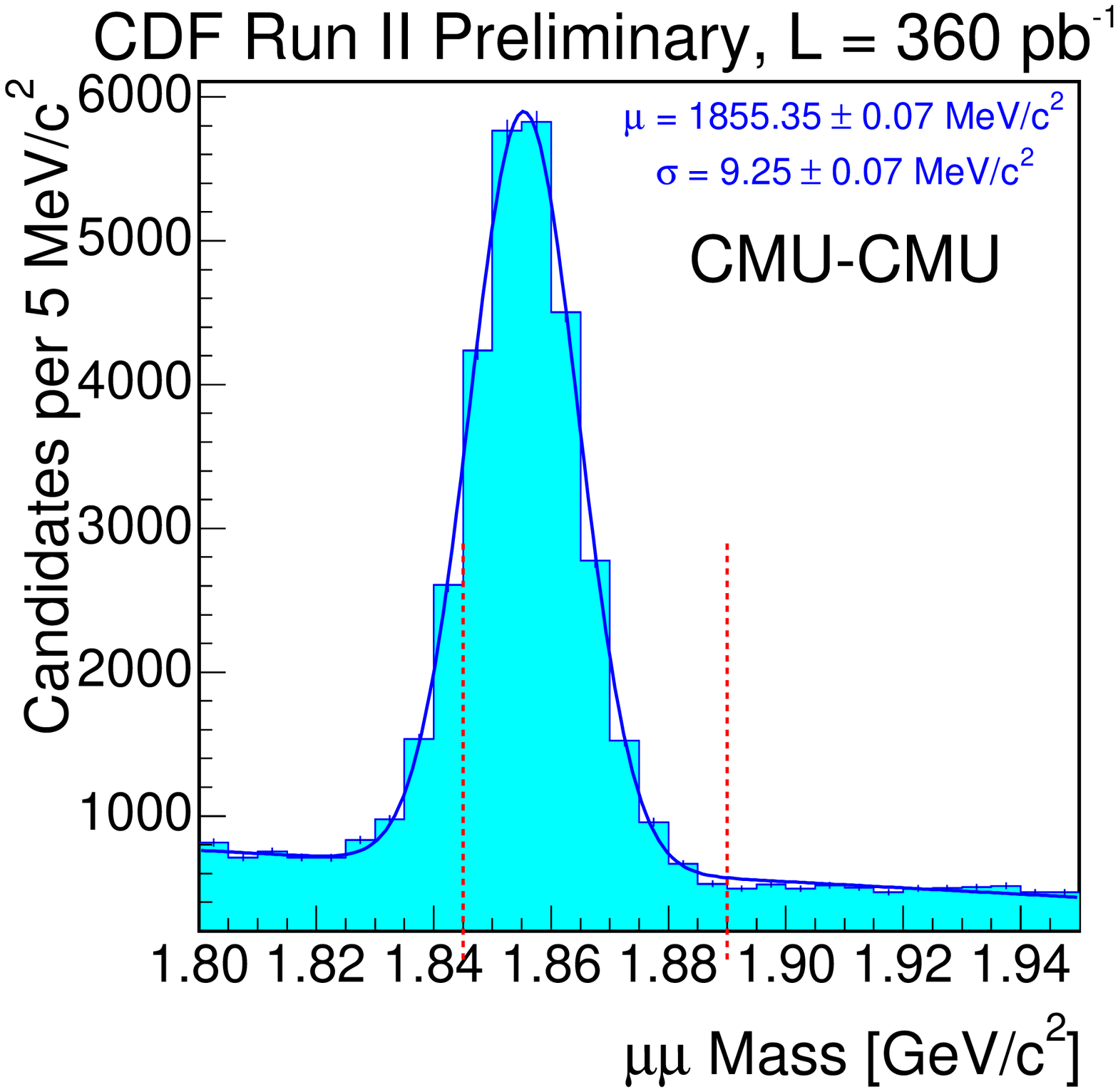}
\includegraphics[width=55mm]{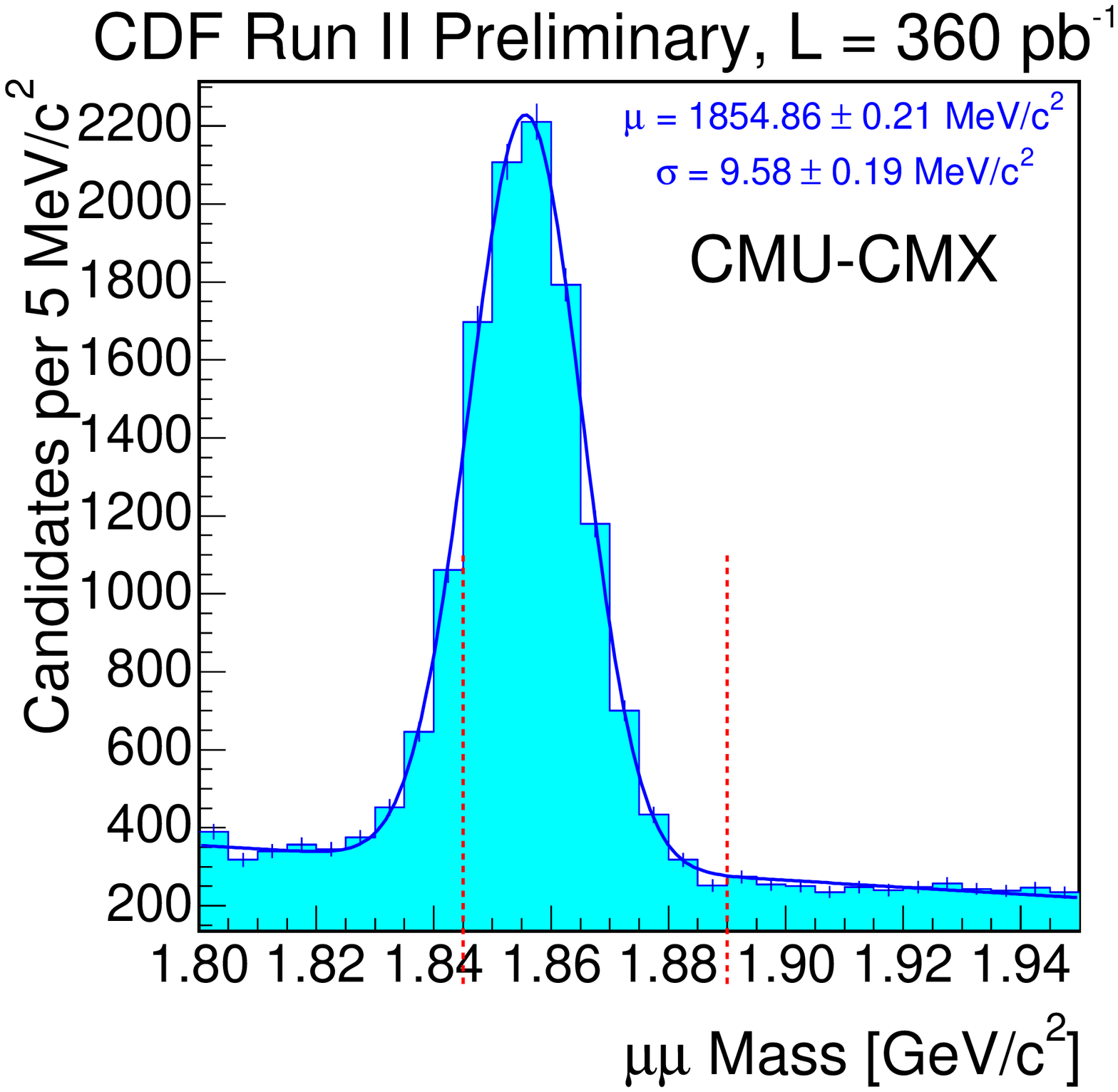}
\includegraphics[width=55mm]{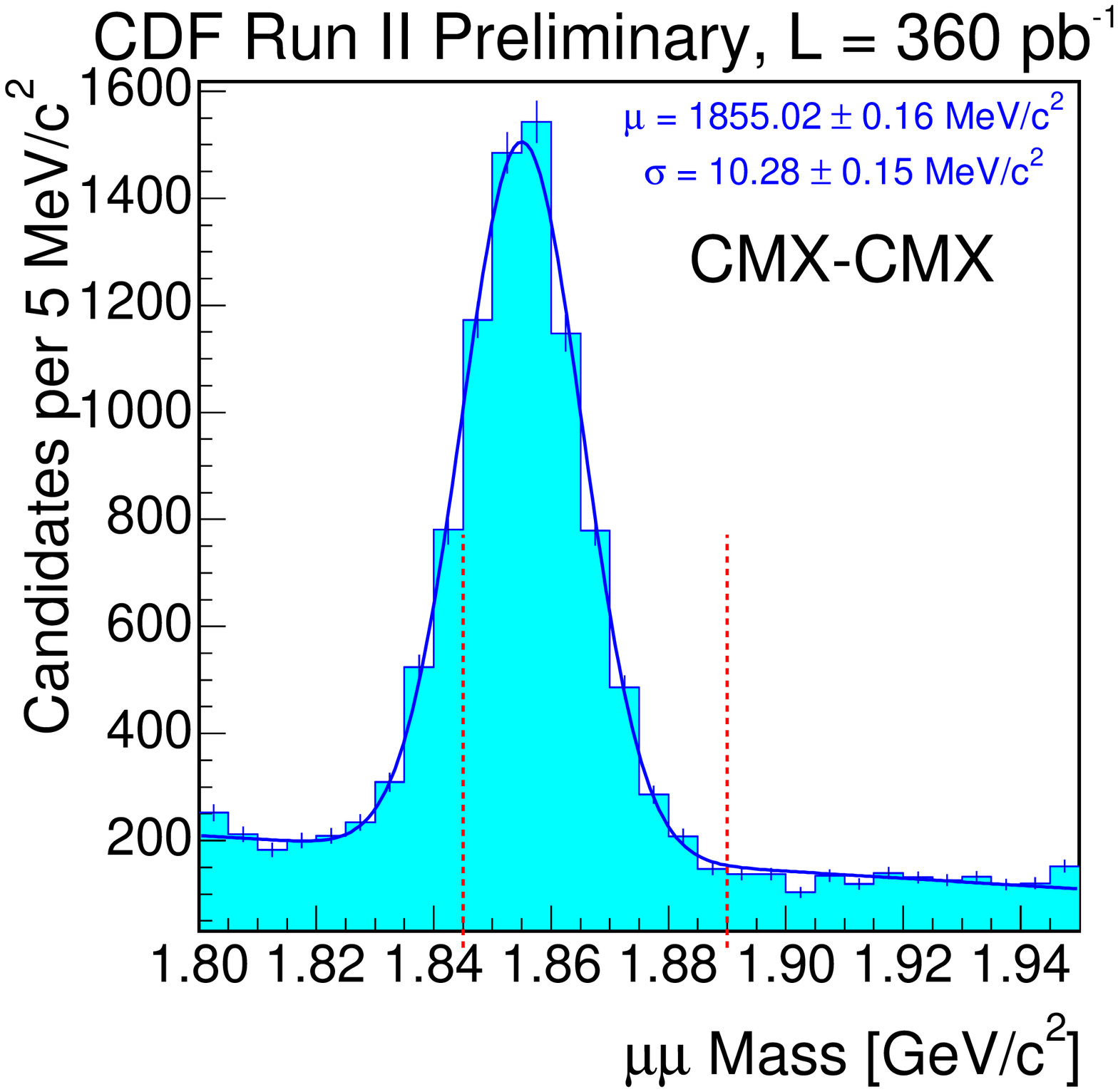}
\caption{The dimuon invariant mass plot before muon selection is applied for candidates where both tracks point to the central muon detector (left), one points to the central and the other to the extension muon detector (center), and both tracks point to the extension muon detector (right).
The large peaks are normalization mode $D^0 \rightarrow \pi^+\pi^-$ decays.
The vertical dotted lines indicate the $\mu\mu$ search window.} 
\label{fig:dmmMass}
\end{figure*}

The $D^0 \rightarrow \mu^+\mu^-$ search is normalized to the Cabibbo suppressed $D^0 \rightarrow \pi^+\pi^-$ decay \cite{dmm}.
Two body $D^0$ decays are reconstructed from the trigger tracks in events selected by the TTT from $360\,\mathrm{pb}^{-1}$ of integrated luminosity.
Combinatorial background to the normalization and search channels is greatly reduced by requiring an associated pion from the decay chain $D^{*+} \rightarrow D^0\pi^+$.
The small mass difference between pions and muons leads to invariant mass distributions that overlap significantly as shown in Fig.~\ref{fig:dmmMass}.
We identify a track as a muon based on hits in either the central muon detector (CMU) or the muon extension (CMX), as determined by projecting the track from the central outer tracker to the muon detectors.

A careful study of background sources revealed that a significant background exists from pairs of real muons coming from the sequential decay chain $B \rightarrow D\mu X$ and $D \rightarrow \mu Y$.
This background is suppressed using a likelihood ratio based on decay length, decay length significance, and impact parameter of the candidate $D^0$ to the primary vertex.

When the data is unblinded we observe three central-central events, zero central-extension events, and one extension-extension event, consistent with the background estimate.
The derived $90\%$ and $95\%$ Bayesian credibility limits are displayed in Table~\ref{limits}.

\section{RESULTS}

\begin{figure*}[htbp]
\centering
\includegraphics[width=85mm]{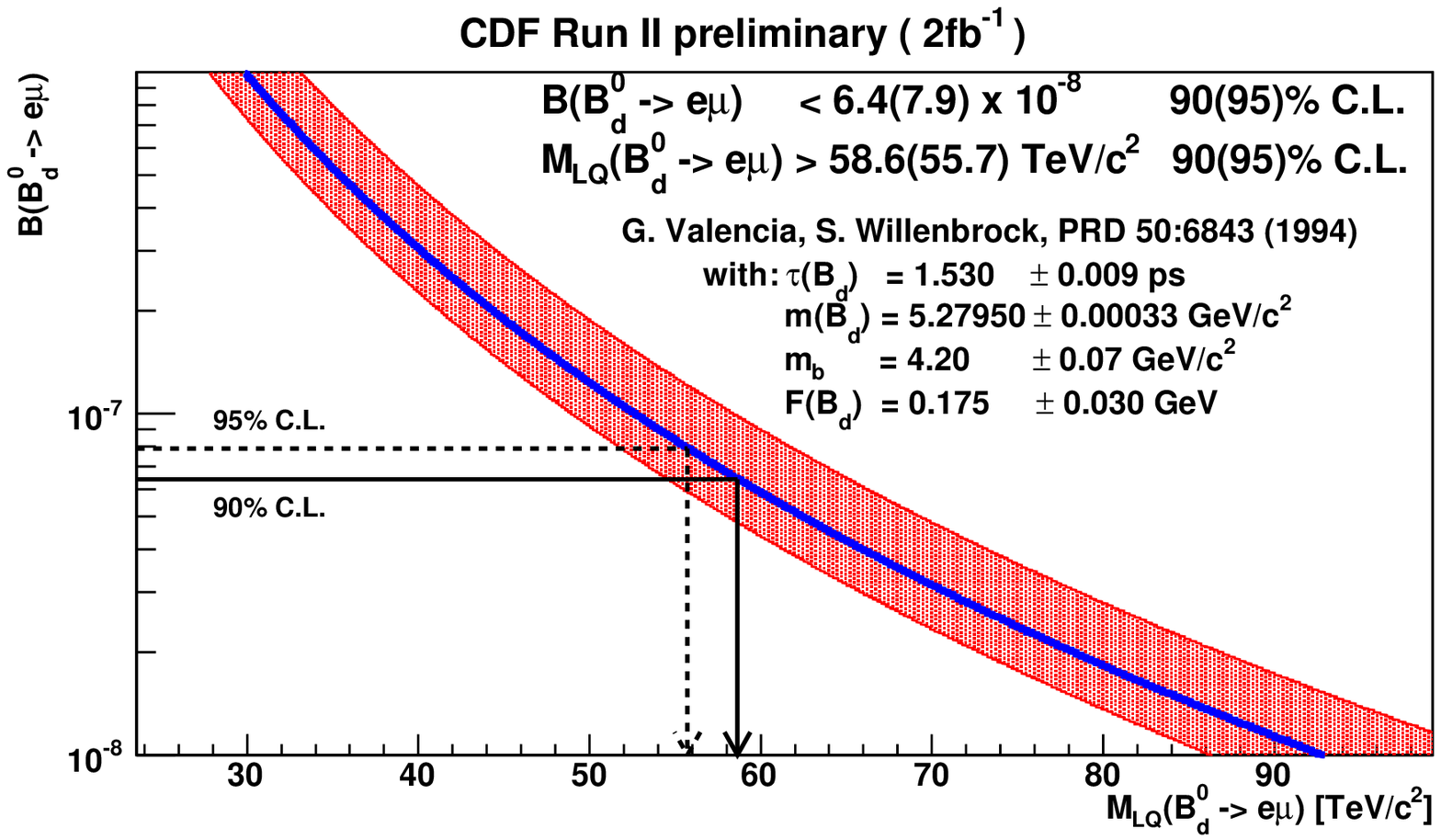}
\includegraphics[width=85mm]{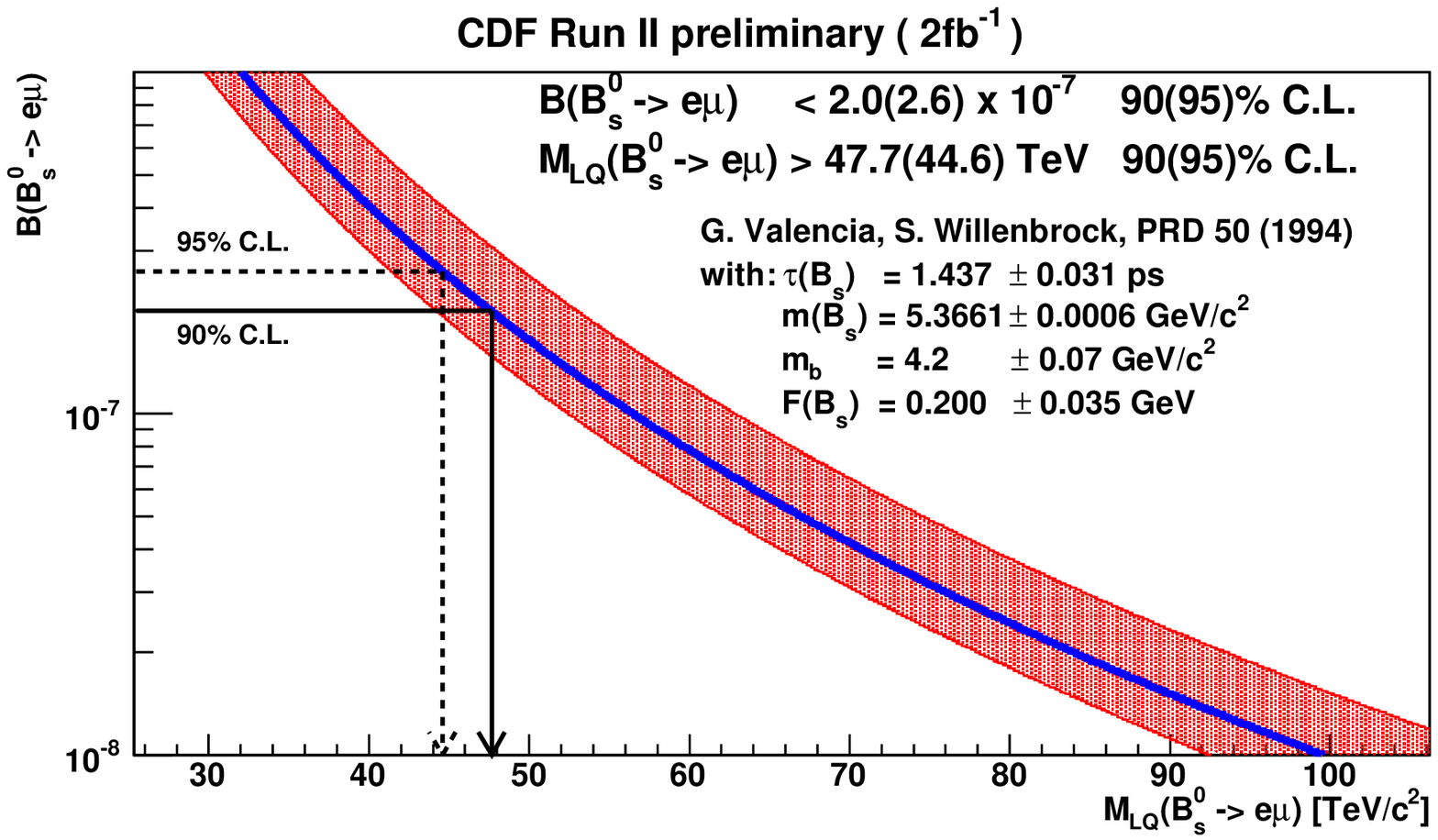}
\caption{Pati-Salam leptoquark mass limits derived from the $B^0_d \rightarrow e\mu$ (left) and $B^0_s \rightarrow e\mu$ (right) branching fraction limits.} \label{fig:LQmasslimits}
\end{figure*}

The limits from these analyses are displayed in Table~\ref{limits}, along with the previous best published limit and the experiment that produced it.
From these limits we derive limits on NP models.
The $B^0_s \rightarrow \mu^+\mu^-$ limit is used to restrict the allowed parameter space for supersymmetry models.
The $B^0_{(s)} \rightarrow e^{\pm}\mu^{\mp}$ limits have been used to derive mass limits on Pati-Salam leptoquarks \cite{psleptoquarks} (Fig.~\ref{fig:LQmasslimits}).
The mass limits are in the range of $50\,\mathrm{TeV}/c^2$ for both decays.

\begin{table}[htbp]
\begin{center}
\caption{Summary of the rare decay branching fraction results.
The values are 90\% (95\%) Bayesian credibility limits multiplied by $10^8$.}
\begin{tabular}{|l|c|c|c|}
\hline \textbf{Mode} & \textbf{CDF} & \textbf{Previous best} &
\textbf{Experiment}
\\ \hline 
$B^0_s \rightarrow \mu^+\mu^-$ & 4.7 (5.8) & 9.4 &  D\O \\
$B^0_d \rightarrow \mu^+\mu^-$ & 1.5 (1.8) & 3.9 & CDF \\
$B^0_s \rightarrow e^{\pm}\mu^{\mp}$ & 20 (26) & 610 & CDF \\
$B^0_d \rightarrow e^{\pm}\mu^{\mp}$ & 6.4 (7.9) & 9.2 & BABAR \\
$B^0_s \rightarrow e^+e^-$ & 28 (37) & 5400 &  L3\\
$B^0_d \rightarrow e^+e^-$ & 8.3 (10.6) & 11.3 & BABAR \\
$D^0 \rightarrow \mu^+\mu^-$ & 43 (53) & 130 & BABAR \\
\hline
\end{tabular}
\label{limits}
\end{center}
\end{table}

The $D^0 \rightarrow \mu^+\mu^-$ limit has been used to set a limit on the product of normalized coupling constants in R-parity violating SUSY \cite{bghp}:
\begin{equation}
\widetilde{\lambda}'_{21k} \widetilde{\lambda}'_{22k} \leq 6\sqrt{\mathcal{B}(D^0 \rightarrow \mu^+\mu^-)} = 3 \times 10^{-3}
\end{equation}

\begin{acknowledgments}
Work supported by Department of Energy contract DE-AC02-76SF00515.
\end{acknowledgments}

\end{document}